\documentclass[aps,prb,twocolumn,groupedaddress,floatfix]{revtex4}
\usepackage{graphics,graphicx}
\usepackage{color}
\usepackage{amssymb,amsmath}
\usepackage{bm}

\newcommand{\req}[1]{Eq.~(\ref{#1})}
\renewcommand{\eqref}[1]{Eq.~(\ref{#1})}
\newcommand{\reqs}[1]{Eqs.~(\ref{#1})}
\newcommand{\rref}[1]{(\ref{#1})}

\newcommand{\Del}{{\mbox{\footnotesize $\Delta$}}}

\newcommand{\be}{\begin{equation}}
\newcommand{\ee}{\end{equation}}
\newcommand{\bea}{\begin{eqnarray}}
\newcommand{\eea}{\end{eqnarray}}
\newcommand{\wm}{\omega_n}
\newcommand{\Tso}{T_{s,0}}
\newcommand{\Tco}{T_{c,0}}
\newcommand{\Gt}{\Gamma_t}
\newcommand{\Gp}{\Gamma_\pi}
\newcommand{\gm}{g_{\omega_n}}

\newcommand{\gmM}{g_{\omega_n}(M)}
\newcommand{\fm}{f_{\omega_n}}
\newcommand{\Sm}{S_{\omega_n}}

\begin{tiny}
\end{tiny}

\begin{document}
\unitlength = 1mm
\title{Phase diagram of iron-pnictides if doping acts as a disorder}
\author{M.~G.~Vavilov and  A.~V.~Chubukov}
\affiliation{~Department of Physics,
             University of Wisconsin, Madison, Wisconsin 53706, USA}
\date{October 5, 2011}

\begin{abstract}
We obtain and analyze the phase diagram of doped iron-pnictides under the assumption that doping adds non-magnetic
 impurities to the system but does not change the densities of carriers. We show that the phase diagram is quite similar to the one obtained under the opposite rigid band assumption. In both cases, there is a phase where $s^{\pm}$ superconductivity and
antiferromagnetism  co-exist. We evaluate the jump of the specific heat, $\Del C$, at the superconducting $T_c$ across the phase diagram and show that $\Del C/T_c$ is non-monotonic, with the maximum at the onset of the co-existence phase. Our results are in quantitative agreement with experiments on some iron-pnictides.
\end{abstract}
\maketitle

\section{Introduction}
How chemical doping of iron-pnictides affects their electronic
structure is not fully understood yet and is subject of debates.
In most studies it is assumed that doping  does not affect the
rigid band picture and only changes the densities of holes and
electrons~\cite{convent}.  An alternative scenario\cite{Sawatzky} is that doping
 does not affect the carrier density
but rather
  introduces non-magnetic impurities and hence increases disorder.
  Angle-resolved photoemission (ARPES) experiments on  122 materials Ba(Fe$_{1-x}$Co$_x$)$_2$As$_2$ and Ba$_{1-x}$K$_x$Fe$_2$As$_2$ are usually interpreted in favor of the rigid band scenario.
Within this scenario, if magnetism prevails at zero doping, the system moves from a spin-density-wave (SDW) phase to a superconducting (SC) state, and for some model parameters there is a mixed phase, where SDW and SC orders co-exist.\cite{fernandes,anton,zlatko}
Recent ARPES experiments on Ru--doped BaFe$_2$As$_2$, however,
 found~\cite{brouet,adam} that substitution of Fe with Ru
  practically does not change the Fermi surface (FS), yet the phase diagram is
   quite similar to that in other doped 122
  materials:
  as Ru concentration increases, the system moves from an SDW phase to an SC  phase. In between,   there is a region where
   where both SDW and SC orders co-exist, although microscopic  co-existence (as opposed to phase separation)
has not been experimentally proven yet.\cite{adam_private}
   Because FS geometry does not change, it seems natural to assume that  the changes in the phase diagram caused by Ru-substitution are
    predominantly due to dilution and disorder associated with it.
We also note that disorder may be introduced directly to pnictide materials by irradiation.~\cite{tarantini2010,kim2010}

In the present paper, we address the issue of what is  the
phase diagram of a doped 122 Fe-pnictide if
doping does not affect the carrier density but rather
 introduces non-magnetic impurities. We show that the phase diagram
 is actually the same as in the rigid band scenario. Namely,
 as doping increases, first an SDW phase becomes a mixed phase, then the system becomes a pure $s^{\pm}$ SC, and at even larger dopings $s^{\pm}$ SC is
  destroyed by disorder.
  This result may look somewhat counter-intuitive because non-magnetic impurities are pair-breaking for an $s^{\pm}$ SC. It turns out, however, that impurities damage SDW order stronger than they damage $s^{\pm}$ SC
because both intra  and inter-band impurity scattering is destructive for SDW,~\cite{kulikov1984} while only inter-band scattering is pair-breaking for an $s^{\pm}$ SC.\cite{dolgov2009,VVC2,bang}
 Because of this disparity,
 the actual magnetic $T_s$ becomes smaller than the superconducting $T_c$ when the density of impurities exceeds a certain threshold, even for the undoped case $T_s > T_c$.
 There is no a'priori guarantee that a mixed state emerges near the point where $T_s = T_c$, \textit{i.e.} a first order transition from an SDW to a SC is another option. Our
  calculation shows that the mixed state does appear, see Fig. \ref{fig:1}a.
 For such a phase diagram to emerge, the magnetic SDW $T_{s,0}$ for undoped material should not be too strong compared to $T_{c,0}$, see below.  If $T_{s,0}/T_{c,0}$ is too large, $T_s$ remains higher than $T_c$ down to $T_c \to 0$, even though $T_s$ decreases faster.

There is another reason to analyze the phase diagram
 assuming that doping introduces disorder.
The measurements of the specific heat jump $\Del C$ at $T_c$ across the phase diagram have demonstrated~\cite{budko_09,hardy_10,hardy_10a,large} that $\Del C/T_c$ is non-monotonic and  has a maximum at optimal doping that almost coincides with the onset of the co-existence phase.
The slopes of $\Del C/T_c$  are similar, although not exactly identical,
  upon deviations from optimal doping into both directions.
 This similarity raised speculations that the behavior of $\Del C/T_c$ in underdoped and overdoped regimes
 may be
  related.
 Within
 the rigid band model, $\Del C/T_c$ has a peak at the onset of a mixed phase and rapidly decreases at lower doping.~\cite{vvc_rigid} However, the
   reduction of $\Del C/T_c$ at higher doping cannot be
 straightforwardly
    explained
  within the rigid band model.

In the disorder model,
 the behavior of $\Del C/T_c$ across the whole phase diagram is determined by a single parameter, the density of impurities $n_{\rm imp}$, and the forms of $\Del C/T_c$
  in the underdoped and overdoped regimes are related.
We find that
 in the disordered model, $\Del C/T_c$ indeed decreases on both sides of optimal doping, as shown in
  Figs. \ref{fig:1}b and \ref{fig:2}.
  The specific heat is discontinuous at the onset of the mixed phase within the mean field approximation,
  but becomes rounded once fluctuations are taken into account.
  The decrease of $\Del C/T_c$ away from the maximum shows rather similar,  although  not identical,
   behavior in under- and over-doped regimes, with roughly quadratic dependence on the transition temperature $T_c$, see Fig.~\ref{fig:2}b.
This behavior is in {\it quantitative} agreement with experiments.\cite{budko_09,hardy_10,hardy_10a}

The fact that the phase diagram and the behavior of $\Del C/T_c$ are similar in the rigid band and the disorder
models is encouraging, since
 the two models are complementary to  each other.
In general,
 a chemical doping acts in both ways: (1) doping
  introduces some extra carriers and (2) increases  impurity density.
The relative magnitude of the two effects depends on materials.
We argue in this regard that quite similar behavior observed in Ru, Co, and K - doped BaFe$_2$As$_2$~\cite{budko_09,hardy_10,hardy_10a,large} is not a coincidence but rather a quite generic feature of iron-pnictides.

The paper is organized as follows. In the next section we discuss the model and introduce SDW and SC order parameters and describe the formalism used for calculations.
 In Sec. \ref{sec:3} we analyze the phase diagram as a function of impurity concentration, by solving linearized gap equation for one order parameter, SDW or SC, when the second parameter is either absent or present.
 Section~\ref{sec:4} presents calculations of the superconducting order parameter near the transition to a superconducting state.  In Sec.~\ref{sec:5}, we consider specific heat jump at the onset of superconductivity.
We provide our conclusions in Sec. \ref{sec:6}.

\section{The model }
\label{sec:2}

\subsection{General formulation}
Our goal is to demonstrate that the phase diagram remains the same if we associate doping with disorder rather than with the changes to the FS in the rigid band picture.
We adopt the same minimal model that was used in earlier works within the rigid band approach.\cite{vvc_rigid} Namely, we consider  a two band metal with cylindrical FSs for electron and hole-type excitations.  The cylindrical FSs have circular cross-sections
of equal radii centered at $(0,0)$ with a hole--like dispersion and $\bm{Q} = (0,\pi)$ with an electron--like dispersion.
The free fermion part of the Hamiltonian in this case of perfect nesting is
represented by
\begin{equation}
{\cal H}_0=-\sum_{\bm{p},\alpha}\xi(\bm{p})\hat c_{\bm{p}\alpha}^\dagger \hat c_{\bm{p}\alpha}^{} +
\sum_{{\bm{p}},\alpha}\xi(\tilde{\bm{p}})
\hat f_{\tilde{\bm{p}}\alpha}^\dagger
\hat f_{\tilde{\bm{p}}\alpha}^{},
\nonumber
\end{equation}
where operators $\hat c$ annihilate hole--like fermions near $(0,0)$ and operators $\hat f$ annihilate  electron--like fermions near $\bm{Q}$.  The fermionic
 dispersion
 is given by $\xi(\bm{p})=\bm{p}^2/2m-\mu$, and the 
momentum $\tilde{\bm{p}}$ of electron excitations is measured as a deviation from $\bm{Q}$, $\tilde{
\bm{p}}=\bm{p}-\bm{Q}$.

 We consider an effective low-energy theory with
 the high-energy cutoff $\Lambda$ and  angle-independent interactions in the SDW channel and in the $s^\pm$ SC channel with the couplings $\lambda_{\rm sdw}$ and $\lambda_{\rm sc}$.\cite{anton,CEE,saurabh2010,Wang2009,thomale2009,platt2009}
We treat these  interactions within a mean field approximation, by introducing SC  and SDW  order parameters,
 $\Delta$ and ${\bf M}$, respectively,  and decomposing  the four-fermion interactions into effective quadratic terms with
  $\Delta$ and ${\bf M}$ in the  prefactors.
The full   mean-field Hamiltonian is quadratic in fermionic operators and can be written as
\begin{equation}
{\cal{H}}=\frac{1}{2}\sum_{\bm{p},\alpha,\beta}
\overline{\Psi}_{\bm{p},\alpha}\hat{H}_{\bm{p},\alpha,\beta}
\Psi_{\bm{p},\beta},
\end{equation}
where $\overline{\Psi}_{\bm{p},\alpha}=(\hat c^\dagger _{\bm{p},\alpha},\ \hat c^{} _{-\bm{p},\alpha},\ \hat f^\dagger_{\bm{p},\alpha},\ \hat f^{}_{-\bm{p},\alpha})$ and $\Psi_{\bm{p},\alpha}$ is a conjugated column. The Hamiltonian matrix $\hat{H}_{\bm{p},\alpha,\beta}$ can be written in the form\cite{anton}
\begin{equation}
\begin{split}
\hat H & = \hat H_0+\hat H_{\rm mf}; \quad \hat H_0=-\xi\hat{\tau}_3\hat{\rho}_3\hat{\sigma}_0\\
\hat H_{\rm mf} &=
-\Delta\hat{\tau_2}\hat{\rho}_3\hat{\sigma}_2
+\hat{\tau}_3\hat{\rho}_1(\bm{M}\hat{\bm{\sigma}}).
\end{split}
\end{equation}
Here,
the Pauli matrices $\hat{\tau}_i$, $\hat{\rho}_i$ and $\hat{\sigma}_i$ are defined in the Gorkov-Nambu, band, and spin spaces,
 respectively, where $i=0,1,2,3$ and
 matrices with $i=0$ are unit matrices.
A fermion Green's function $\hat G(\wm,\bm{p})$ is defined as a solution to
 \begin{subequations}
\begin{equation}
\left(i\wm-\hat H_{\bm{p}}-\hat{\Sigma} \right)\hat G(\wm,\bm{p}) = \hat 1, \label{eq:Gl}
\end{equation}
and the conjugated equation is
\begin{equation}
\hat G(\wm,\bm{p}) \left(i\wm-\hat H_{\bm{p}}-\hat{\Sigma} \right) = \hat 1 \label{eq:Gr}.
\end{equation}
\end{subequations}
Here $\hat{\Sigma}$ is the self energy for scattering off disorder and  $\wm=2\pi T_m (n+1/2)$ with integer $n$ are Matsubara frequencies.

We describe disorder scattering within  the Born approximation and assume that the Born scattering amplitude $U(\bm{q})$  is characterized by a constant $U_0$ for scattering within the same band and $U_\pi$ for scattering between the two bands.\cite{dolgov2009,VVC2,bang}  In this
  approximation, the self-energy  is
\begin{equation}
\begin{split}
\hat{\Sigma}(\wm)= &  \frac{4\Gamma_0}{\pi N_F}\int \frac{ d\bm{p}}{(2\pi\hbar)^2} \hat{\tau}_3\hat{\rho}_0 \hat\sigma_0 \hat{G}(\wm,\xi) \hat{\tau}_3\hat{\rho}_0\hat\sigma_0\\
+& \frac{4\Gp}{\pi N_F}\int \frac{ d\bm{p}}{(2\pi\hbar)^2} \hat{\tau}_3\hat{\rho}_1 \hat\sigma_0\hat{G}(\wm,\xi) \hat{\tau}_3\hat{\rho}_1\hat\sigma_0 .
\label{eq:Sigma}
\end{split}
\end{equation}
  where we  introduced  disorder  scattering rates
\begin{equation}
\Gamma_0=\frac{\pi N_F n_{\rm imp}}{4} |U_0|^2,
\quad
\Gp=\frac{\pi N_F n_{\rm imp}}{4}  |U_\pi|^2.\label{eq:rates}
\end{equation}
$\Gamma_0$ characterizes the rate of electron collisions with impurities  in which  the electron remains in its original band, while $\Gp$ is the rate of collisions resulting in electron
  transfer
 between the two bands.  $N_F$ in Eq.~(\ref{eq:rates}) is the total quasiparticle density of states (DoS) at the Fermi energy
 ( the DoS per spin per band is $N_F/4$).
  We assume that only  the impurity density $n_{\rm imp}$ changes with doping, i.e., the ratio $\Gp/\Gamma_0$ is doping independent.

The two mean-field parameters $\Delta$ and ${\bf M}$ are obtained self-consistently via the matrix Green's function as
\begin{equation}
\label{eq:scDelta}
\frac{\Delta}{\lambda_{\rm sc}}=\frac{T}{2}\sum_{\wm}\int\frac{ d\bm{p}}{(2\pi\hbar)^2}
{\rm Tr}\left\{\hat G(\wm,\bm{p})  
\hat{\tau}^+\left(\hat{\rho}_0+\hat{\rho}_3  \right)\hat{\sigma}^+\right\},
\end{equation}
and
\begin{equation}
\label{eq:scM}
\frac{\bm{M}}{\lambda_{\rm sdw}} =
\frac{T}{4}\sum_{\wm}\int\int \frac{ d\bm{p}}{(2\pi\hbar)^2}
{\rm Tr}\left\{\hat G(\wm,\bm{p})   
\left(\hat{\tau}_0+\hat{\tau}_3\right)\hat{\rho}^+ \hat{\bm{\sigma}}\right\},
\end{equation}
where $\hat{A}^+=(\hat{A}_1+i\hat{A}_2)/2$ for $\hat{A}\to \hat{\rho},\hat{\tau},\hat{\sigma}$.

For  the pure SDW and the pure $s^{+-}$ SC state in the absence of disorder, the solution of the linearized gap equations yield transition temperatures $\Tso=1.13 \Lambda\exp(-2/(N_F\lambda_{\rm sdw}))$ and $\Tco=1.13 \Lambda\exp(-2/(N_F\lambda_{\rm sc}))$.
We consider $\Tso>\Tco$, so that without disorder the SDW phase develops at a higher temperature.

\subsection{Eilenberger equation}

To treat superconductivity and magnetism in the presence of disorder, it is convenient to introduce
 the Eilenberger's Green function
\begin{equation}
\label{eq:Geilenberger}
\hat{\cal{G}}(\wm)=\frac{4i}{\pi N_F}\int \frac{ d\bm{p}}{(2\pi\hbar)^2}  \hat{\tau}_3\hat{\rho}_3 \hat \sigma_0 \hat{G}(\wm,\bm{p})
\end{equation}
which appears both in the self-consistency equations, \reqs{eq:scDelta} and \rref{eq:scM}, and in the expression for the impurity self-energy, \eqref{eq:Sigma}. In particular,  the impurity self energy is
\begin{equation}
\begin{split}
\hat \Sigma & = -i \Gamma_0 \hat\tau_0  \hat\rho_3  \hat\sigma_0  \hat{\cal G} \hat\tau_3  \hat\rho_0  \hat\sigma_0    -i  \Gp \hat\tau_0  (-i\hat\rho_2)  \hat\sigma_0  \hat{\cal G} \hat\tau_3  \hat\rho_1  \hat\sigma_0.
\end{split}
\end{equation}

 To derive the equation for $\hat{\cal{G}}$,
we multiply \req{eq:Gl} by $\hat{\tau}_3\hat{\rho}_3$ from left and subtract \req{eq:Gr}, multiplied by $\hat{\tau}_3\hat{\rho}_3$
from right.  We then multiply the resulting equation by $\hat{\tau}_3\hat{\rho}_3$ from left again.
As a result, the $\hat H_0(\bm{p})$ term falls out.  We
 integrate
 the resulting equation over $\bm{p}$ and obtain the
 equation for $\hat{\cal{G}}(\wm)$ in the form of a commutator:
\begin{equation}
\left[i\wm\hat{\tau}_3\hat{\rho}_3\hat\sigma_0-(\hat H_{\rm mf}+\hat \Sigma)\hat{\tau}_3\hat{\rho}_3\hat\sigma_0;\hat{\cal{G}}(\wm)
\right]=0.
\label{eq:eilenberger}
\end{equation}
This equation is the
 Eilenberger equation,\cite{Eilenberger68,Moor2011} obtained for a
 two-band metal with
 homogeneous in space
SDW and SC order parameters.
 The Eilenberger  equation is consistent with the normalization relations for $\hat{\cal{G}}$: $\rm{Tr}\hat{\cal{G}}(\wm)=0$ and
$\hat{\cal{G}}(\wm)\hat{\cal{G}}(\wm)=\hat 1$.

Without loss of generality, we direct $\bm{M}$ along $z$--axis and parametrize 
 the matrix $\hat{\cal{G}}(\wm)$
 by the  three functions $\gm$, $\fm$ and $\Sm$ as
\begin{equation}
\hat{\cal{G}}(\wm)=\gm\hat{\tau}_3\hat{\rho}_3\hat\sigma_0
+\fm  \hat{\tau}_1\hat{\rho}_0(-i\hat\sigma_2)
+\Sm\hat{\tau}_0(-i\hat{\rho}_2)\hat\sigma_3.
\label{eq:Gparameterization}
\end{equation}
 The function $g_{\omega_n}$ is the normal component of the Eilenberger Green's function, while
the functions $S_{\omega_n}$ and $f_{\omega_n}$ are the two anomalous components, associated with the SDW and SC orders, respectively.

For the above parametrization of $\hat{\cal G}(\wm)$, \eqref{eq:Gparameterization}, the normalization condition $\hat{\cal{G}}\hat{\cal{G}}=1$ reduces to
\begin{equation}
\gm^2-\Sm^2-\fm^2=1, \label{eq:norm}
\end{equation}
  and the commutation relation \req{eq:eilenberger} gives
\begin{subequations}
\begin{eqnarray}
&& i\Delta\gm=\fm(\wm+2\Gp\gm), \label{eq:SCeq}\\
&& iM\gm=\Sm(\wm+2\Gt\gm). \label{eq:SDWeq}
\end{eqnarray}
\end{subequations}
where $\Gamma_t = \Gamma_0 + \Gamma_\pi$.

The self-consistency equations for SDW and SC order parameters, Eqs.  (\ref{eq:scDelta}) and (\ref{eq:scM}), can be rewritten in terms of anomalous SDW and SC components of the Eilenberger's Green function as
\begin{subequations}\label{eq:selfcons_all}
\begin{eqnarray}
\frac{2M}{N_F\lambda_{\rm sdw}}  & =  & -i~2\pi T\sum_{\wm>0}^\Lambda \Sm, \quad
\label{eq:selfcons_sdw}
\\
\frac{2\Delta}{N_F\lambda_{\rm sc}} & = & -i~2\pi T\sum_{\wm>0}^\Lambda \fm.
\label{eq:selfcons_sc}
\end{eqnarray}
\end{subequations}

\section{Phase diagram}\label{sec:3}

We first consider pure SDW and SC states.
 For a pure SDW state we set  $\Delta=0$ and $\fm=0$ in Eqs.~(\ref{eq:norm}) and (\ref{eq:SCeq}), linearize \eqref{eq:SDWeq} in $M$ and find from \eqref{eq:selfcons_sdw} that the  SDW transition temperature $T_s$ evolves with doping as
\begin{equation}
\frac{2}{N_F\lambda_{\rm sdw}}=2\pi T_s \sum_{ n \geq 0}^{\Lambda/2\pi T_s}
\frac{1}{\pi T_s (2n+1)+ 2 \Gt}.
\label{n_2sdw}
\end{equation}
This equation can be rewritten in terms of the transition temperature $\Tso$ to SDW phase at $\Gt=0$
 as
 \begin{equation}
\ln\frac{\Tso}{T_s}=
\psi\left(\frac{1}{2}+\frac{\Gamma_t}{\pi T_s}\right)-\psi\left(\frac{1}{2}\right),
\label{eq:Ts}
\end{equation}
where
 $\psi (x)$ is the digamma function.\cite{kulikov1984}

For a pure SC state we set  $M=0$ and $\Sm=0$  in
 Eqs. (\ref{eq:norm}) and  (\ref{eq:SDWeq})
and linearize \eqref{eq:SCeq} in $\Delta$. We obtain from \eqref{eq:selfcons_sc}
\begin{equation}
\frac{2}{N_F\lambda_{\rm sc}}=2\pi T_c \sum_{ n \geq 0}^{\Lambda/2\pi T_c}\frac{1}{\pi T_c (2n+1)+ 2 \Gp}.
\label{n_2}
\end{equation}
Re-expressing the result in terms of the superconducting transition temperature  $T_{c,0}$  in a clean system and without SDW, we re-write \eqref{n_2} as
\begin{equation}
\ln\frac{\Tco}{T_c}=\psi\left(\frac{1}{2}+\frac{\Gp}{\pi T_c}\right)-\psi\left(\frac{1}{2}\right),
\label{eq:Tc}
\end{equation}
which is similar to the equation for $T_c$ in conventional $s-$wave superconductors with magnetic impurities\cite{AG-spins} and in unconventional $d-$wave superconductors with potential impurities.\cite{Balatsky2006,Galitski2008,Vorontsov2010}
Note that only inter-band scattering
 $\Gp$,
is pair-breaking for $s^{\pm}$ SC.

\begin{figure}
\centerline{\includegraphics[width=0.95\linewidth]{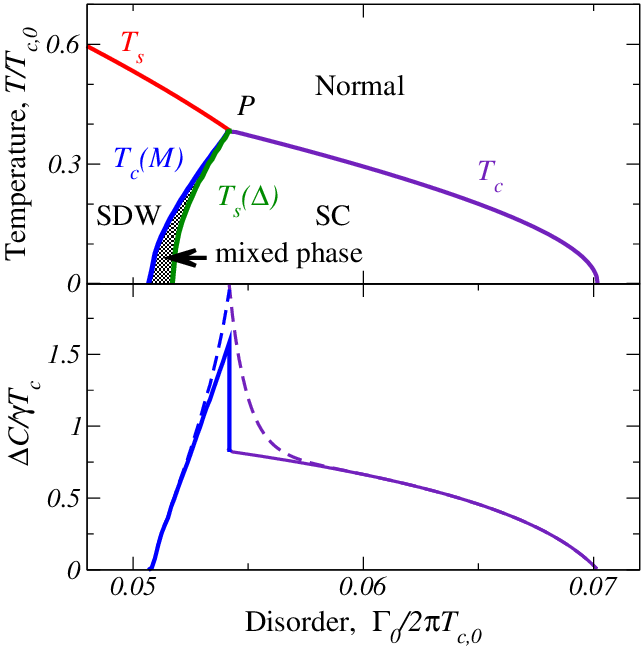}}
\caption{(Color online)
 \textit{Upper panel}: The phase diagram as a function of disorder, measured in units $\Gamma_0/2\pi T_{c,0}\propto n_{\rm imp}$, for on-site disorder ($\Gp=\Gamma_0$), and $\Tso/\Tco=1.7$.
The four transition lines terminate at a
 tetra-critical point $P$, where normal, pure SDW, pure SC, and mixed phase meet.
The shaded region represents the mixed phase.  \textit{Lower panel}:
Specific heat jump $\Del C/T_c$ as a function of doping.  Solid curves represent the mean field result
and  the dashed curve
 illustrates the effect of thermodynamic fluctuations beyond the mean field theory.}
\label{fig:1}
\end{figure}

Even if $\Tso>\Tco$,  $T_s$ decreases faster than $T_c$ with increasing $n_{\rm imp}$, and at  certain doping the two transition temperatures
 may cross.
 We denote this temperature as
$T_P$.
The condition that $T_P$ exists, i.e. that $T_c$ and $T_s$ cross before $T_c\to 0$, sets the limits on the ratios $\Tso/\Tco$ and  $\Gp/\Gamma_0$. We find that
 $T_s$ and $T_c$ cross if
$\Tco/\Tso>1/( 1+\Gamma_0/\Gp )$.
For  on-cite disorder potential $\Gp=\Gamma_0$, and
$T_P>0$ exists, i.e. SC phase exists,
if $\Tco/\Tso >1/2$. For longer-range impurity potentials, $\Gamma_0 > \Gp$, and the SC phase develops  even for smaller  $\Tco/\Tso$, see Figs. \ref{fig:1}a and \ref{fig:3}a.

To obtain
 the superconducting transition
 temperature $T_c (M)$
  in the presence of pre-existing magnetism
 one has to solve the linearized equation for $\Delta$ at a finite $M$. Now $g_{\omega_n}$ depends on $M$ (i.e., $g_{\omega_n} = \gmM$), and we
 have from  \eqref{eq:selfcons_sc}
\begin{equation}
\alpha(T_c(M))   =
\frac{2}{ N_F \lambda_{sc}} ,
\label{n_1}
\end{equation}
where
\begin{equation}
\alpha(T) =
 2\pi T  \sum_{\omega_n >0}^\Lambda \frac{\gmM}{\wm+2\Gp\gmM}.
\label{eq:alpha}
\end{equation}
The temperature dependence in the r.h.s. of \eqref{eq:alpha} is via $\omega_n = \pi T (2n+1)$ and also via $\gmM$ because $M$ depends on temperature.
The summation in $\alpha(T)$ has a logarithmic dependence on the high-energy cut-off $\Lambda$.  This
 dependence
  can be eliminated in favor of the transition temperature $T_c$ at $M=0$. Subtracting \eqref{n_2} from \eqref{n_1},
 we obtain after a simple algebra an equation on $T_c (M)$ in the form
\begin{equation}
\begin{split}
{\cal L}(T_c(M),T_c,\Gp)=
\sum_{\wm>0}^\Lambda  
\frac{2 \pi T_c (M)\wm [\gmM-1]}{(\wm+2\Gp)(\wm+2\Gp\gmM)}  ,
\label{n_2_2}
\end{split}
\end{equation}
where
\begin{equation}
\begin{split}
{\cal L}(T_1, T_2,\Gamma)
 = \ln\frac{T_1}{T_2}+\psi\left(\frac{1}{2}+\frac{\Gamma}{\pi T_1}\right)-
\psi\left(\frac{1}{2}+\frac{\Gamma}{\pi T_2}\right).
\end{split}
\label{eq:L}
\end{equation}

We calculate $\gmM$ as a function of temperature at a given impurity concentration. For this purpose, we express $S_{\omega_n} (M) $ in terms of $\gmM$ using \eqref{eq:SDWeq},
\begin{equation}
\Sm(M)= \frac{iM  \gmM }{\wm+2\Gt \gmM },
\label{eq:Sm0}
\end{equation}
substitute the result into \eqref{eq:norm} with $\fm=0$, and obtain the fourth--order algebraic equation for $\gmM$ as a function of $M$:
  \begin{equation}
\gm^2(M) + \frac{M^2 \gm^2(M)}{(\omega_n + 2 \Gamma_t \gmM)^2}=1.
  \label{new_a}
  \end{equation}
We solve this equation, obtain $\gmM$, substitute the result back into Eqs.~(\ref{eq:Sm0}) and 
 (\ref{eq:selfcons_sdw}),  utilize the definition of $T_s$, and obtain the non-linear equation for  $M = M(T)$ in the form
\begin{equation}
{\cal L}(T,T_s,\Gt)=2\pi T\sum_{\wm>0}\frac{\wm \left[\gmM-1\right] }{(\wm+2\Gt)(\wm+2\Gt \gmM)},
\label{eq:MnoSC}
\end{equation}
where $\gmM$ is a solution of \eqref{new_a},
one has to choose the  branch with $g_{\omega_n} (M=0) =1$.
 Solving (\ref{eq:MnoSC}) we obtain $M(T)$, and hence $g_{\omega_n} (T)$. Substituting the  result into (\ref{n_1}) and (\ref{eq:alpha}) we obtain the superconducting transition temperature
$T_c (M)$ in the mixed phase as a function of doping.

We numerically evaluate
 $T_c (M)$
 at different dopings in the mixed phase
  and plot the result  in Figs. \ref{fig:1}a and \ref{fig:3}a.
As the doping decreases from its optimal value, $M$ increases at a given temperature $T$, and $T_c (M)$ rapidly drops.
This is expected since SDW and SC order parameters compete with each other.
At $M \to 0$, $\gmM \to 1$ and Eq. (\ref{n_2_2}) yields
$T_c(M) \to T_c$, as expected.

A similar calculation of the SDW transition temperature from the preexisting SC phase, $T_s (\Delta)$,
  shows that
  $T_s (\Delta)$ decreases as $\Delta$ increases,
   due to the same kind of competition.
   Furthermore,
   the $T_s (\Delta)$ curve actually bends toward smaller dopings, see Figs.~\ref{fig:1}a and \ref{fig:3}a,
 so that
  with decreasing temperature  the system moves
  from a pure SDW
    magnet
    to a pure superconductor through a mixed phase. The bending of the $T_s (\Delta)$ curve is in agreement with the general analysis in Ref.~\onlinecite{moon_sachdev}.
    The four curves $T_c$, $T_c(M)$, $T_s$, and $T_s (\Delta)$
     meet at the tetracritical point $P$, as shown in Figs.~\ref{fig:1}a and \ref{fig:3}a. The corresponding temperature $T_P$
      is the highest superconducting transition temperature.  We also see from numerics that, despite bending, the curve $T_s (\Delta)$ is always located to the right of the curve $T_c (M)$, i.e. if one increases disorder  at a given $T$ or decreases  $T$ at a given disorder, the system with the SDW order first becomes unstable toward
  an intermediate mixed phase where SDW and SC orders co-exist,  and only then SDW order disappears.

The intermediate mixed phase was earlier found in the rigid band model.~\cite{fernandes,anton}  However,  in that model it only appears
at a finite ellipticity of electron pockets, while for circular hole and electron FSs doping gives rise to a first order transition between pure SDW and pure SC phases.  In the disorder model, the mixed phase appears already for circular  hole and electron pockets and by continuity should also exists when electron pockets have weak ellipticity. 
 We, however, did not analyze the whole range of ellipticities and therefore cannot exclude a possibility of a first order transition for strongly elliptical electron FSs.

\section{Superconducting order parameter near $T_c$}\label{sec:4}

We verified that the mixed phase does indeed exist in the disorder model with circular FSs by expanding  in \eqref{eq:SCeq} to order $\Delta^3$ and solving the equation for $\Delta$ in the presence of  $\bm{M}$ at a temperature slightly below  $T_c (M)$.  The expansion yields, quite generally:
\be
\alpha (T) - \beta \Delta^2 = \frac{2}{\lambda_{\rm sc} N_F}
\label{eq:selfcons1}
\ee
where $\alpha(T)$ is introduced in \eqref{eq:alpha}
 and
 $\beta = \beta (T_c (M)) $ is  given by \eqref{eq:beta} below.
 Near $T_c(M)$, we have
$\alpha(T) = \alpha (T_c (M)) + \alpha^{\prime} (T_c (M))  \left(T - T_c (M)\right)$ and $\alpha (T_c (M))$, see \eqref{n_1}.
On general grounds, $\alpha^\prime (T_c (M))$  must be negative
for a SC phase to develop as $T$ decreases, and we indeed show below that $\alpha^\prime (T_c (M)) <0$.
The type of the transition is, however, determined by the sign of $\beta$. The mixed phase exists if  $\beta>0$ because then $\Delta$ gradually grows as $T$ decreases. If $\beta<0$, $\Delta$ changes discontinuously around $T_c (M)$ and the SDW and SC phases are separated by the first-order transition.~\cite{fernandes,anton}

\begin{figure}
\centerline{\includegraphics[width=0.95\linewidth]{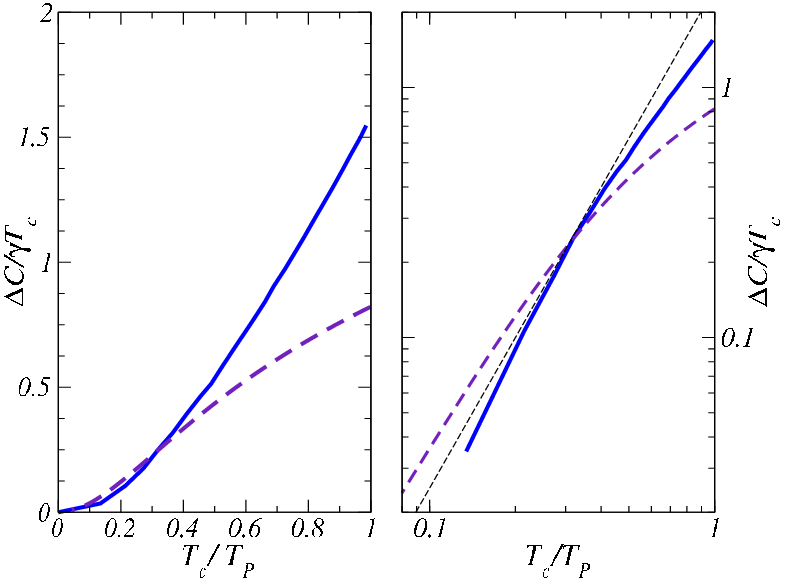}}
\caption{(Color online)
The specific heat jump $\Del C/T_c$ for $\Gp=\Gamma_0$, and 
$\Tso=1.7 \Tco$,
 as a function of $T_c/T_P$, where $T_P$ is the temperature of the tetra-critical point. Left panel: linear scale, right panel:
 log--log scale.
  The solid line represents the specific heat jump $\Del C/T_c$ at
  $T_c (M)$ in the
   mixed phase  in the underdoped regime, and the dashed line  represents $\Del C/T_c$ in the overdoped regime,
   where SDW order is absent.
A thin dash line represents  a quadratic dependence $\Del C/T_c\propto T_c^2$.
 }
\label{fig:2}
\end{figure}

The coefficient  $\alpha (T)$ can be rewritten in the form of  Eqs. (\ref{n_2_2}) and (\ref{eq:L}):
\begin{equation}
\alpha (T) =  \frac{2}{\lambda_{sc} N_F}- Y(T),\quad
\alpha^{'} (T) = -\frac{\partial Y(T)}{\partial T}
\label{eq:alphamixed}
\end{equation}
where
\begin{equation}
\begin{split}
Y(T) &= {\cal L}\left(T,T_c,\Gp\right)   \\
& +\sum_{\wm>0}\frac{2\pi T\wm (1-\gmM) }{[\wm+2\Gp][\wm+2\Gp \gmM]}.
\label{eq:ABC_1}
\end{split}
\end{equation}
For $M=0$ and in the clean limit $\alpha^\prime (T_c) = -1/T_c$. We verified  numerically that $\alpha^\prime (T_c (M))$ remains negative at $M \neq 0$ and in the presence of disorder, as expected.

Calculations of $\beta$ require more
 care as one has to combine terms coming from the apearance of non-zero $\fm$ in \eqref{eq:norm} and from the expansion of the SDW order parameter 
  $M(\Delta)$ to order $\Delta^2$ as
 $M (\Delta)  = M + \delta M^{(2)}$, where $\delta M^{(2)} \propto \Delta^2$.
Similarly, we introduce $\gm=\gmM+\delta \gm^{(2)}$ and $\Sm=\Sm(M)+\delta \Sm^{(2)}$.  Substituting $\gm$ and $\Sm$  into \reqs{eq:norm} and \rref{eq:SDWeq}, we obtain equations for $\delta \gm^{(2)}$ and $\delta \Sm^{(2)}$:
\begin{equation}
\begin{split}
\gmM\delta \gm^{(2)}-\Sm(M)\delta \Sm^{(2)} & =\frac{1}{2}[\fm^{(1)}]^2,\\
\frac{-iM\wm\delta \gm^{(2)}}{(\wm+2\Gt\gmM)^2}+ \delta\Sm^{(2)}& =  \frac{i\delta M^{(2)} \gmM}{\wm+2\Gt\gmM},
\end{split}\label{eq:linsyst}
\end{equation}
where
\begin{equation}
\fm^{(1)}=\frac{i\Delta\gmM}{\wm+2\Gp\gmM}
\label{eq:fm1}
\end{equation}
and $\gmM$ is defined by \eqref{eq:Sm0}.
 Solving \eqref{eq:linsyst} we obtain
 \begin{equation}
\begin{split}
\delta \gm^{(2)}&=-\frac{1}{2}\frac{\gmM(\wm+2\Gt\gmM)^2}{(\wm+2\Gt\gmM)^3+M^2\wm}\\
&\times\left(\Delta^2 \frac{\wm+2\Gt\gmM}{(\wm+2\Gp\gmM^2}+
\frac{2M\delta M^{(2)}}{\wm+2\Gt\gmM}\right)
\end{split}\label{eq:g2}
\end{equation}
and
\begin{equation}
\begin{split}
\delta \Sm^{(2)}&=-\frac{i}{2}\frac{\gmM(\wm+2\Gt\gmM)^2}{(\wm+2\Gt\gmM)^3+M^2\wm}\\
&\times\left( \frac{\Delta^2 M\wm}{(\wm+2\Gp\gmM)^2(\wm+2\Gt\gmM)}
-2\delta M^{(2)}\right).
\end{split}\label{eq:S2}
\end{equation}
We first evaluate $\delta M^{(2)}$ by substituting $\Sm=\Sm^{(0)}+\Sm^{(2)}$ into \eqref{eq:selfcons_sdw}  and eliminating the SDW coupling constant in favor of the SDW transition temperature $T_s$, see \eqref{n_2sdw}.  We obtain
 \begin{equation}
\delta M^{(2)} {\cal L}(T,T_s,\Gt)=-2 \pi T\sum_{\wm>0}\left(i
\delta \Sm^{(2)}+\frac{\delta M^{(2)}}{\wm+2\Gt }
\right).
\end{equation}
Substituting ${\cal L}(T,T_s,\Gt)\approx {\cal L}(T_c(M),T_s,\Gt) $ from \eqref{eq:MnoSC} and $\delta \Sm^{(2)}$ from \eqref{eq:S2} we obtain
\begin{equation}
\delta M^{(2)}=-\frac{\Delta^2}{2M}\frac{C}{B},
\label{eq:deltaM}
\end{equation}
where the coefficients $B$ and $C$,
 together with the term $A$ which we utilize below, are given by
\begin{subequations}\label{eq:ABC}
\begin{align}
A& =  \pi T \sum_{\wm>0} \frac{\gmM\wm(\wm+2\Gt\gmM)^3}
{[\wm+2\Gp\gmM]^4 D},
\label{eq:A}\\
B& = \pi T\sum_{\wm>0} \frac{\gmM\wm}
{[\wm+2\Gt\gmM] D},\\
C& =  \pi T\sum_{\wm>0} \frac{\gmM\wm(\wm+2\Gt\gmM)}
{[\wm+2\Gp\gmM]^2 D},\\
D& =  [\wm+2\Gt\gmM]^3+\wm M^2.\label{eq:D}
\end{align}
\end{subequations}
Substituting $\delta M^{(2)}$ into (\ref{eq:g2}) we obtain $\delta g^{(2)} \propto \Delta^2$.

We next write $\fm$, defined by \eqref{eq:SCeq} to the third order in $\Delta$
 \be
 f_{\omega_m} = \frac{i \Delta \gmM }{\omega_n + 2 \Gamma_\pi \gmM} +
 \frac{i \Delta \omega_n \gm^{(2)} }{(\omega_n + 2 \Gamma_\pi\gmM )^2},
 \label{n_6}
 \ee
 substitute this expression into \eqref{eq:selfcons_sc},
  and obtain \eqref{eq:selfcons1} with
\be
\beta=\left(A-\frac{C^2}{B}\right)
\label{eq:beta}
\ee
 where $A, B$, and $C$ are given by
 \eqref{eq:ABC}.

Evaluating these coefficients,  we find $AB > C^2$, i.e., $\beta$ is positive. This confirms our numerical result that the phase diagram of the disorder model contains the mixed phase where SDW and SC orders co-exist.

Equation (\ref{eq:selfcons1})
 can also be  applied to
 the transition  from a paramagnetic metal into a pure SC state. In this case, $M=0$, and hence $\gm= \gmM = 1$ and
$C^2/B$ term in \eqref{eq:beta} is absent, i.e. $\beta = A$.
  Substituting $\gmM =1$ into \reqs{eq:A} and \rref{eq:D} we obtain
  \begin{subequations}\label{fr_all}
\begin{align}
\alpha'(T_c) & =   - \frac{1}{T_c}
\left(1-\sum_{m=0}^\infty\frac{\Gamma_\pi/\pi T_c}{(m+1/2+\Gamma_\pi/\pi T_c)^2}
\right), \\
\beta & =  \frac{1}{8\pi^2T^2_c}
\sum_{m=0}^\infty\frac{m+1/2}{(m+1/2+\Gamma_\pi/\pi T_c)^4},
\label{fr_2}
\end{align}
\end{subequations}
where $T_c$ is given by  \req{eq:Tc}.
 Note that, again,  $\alpha'(T_c) <0$ and $\beta >0$.

\section{Specific heat jump at the onset of superconductivity \label{sec:5}}

The specific heat jump at $T_c$ and $T_c (M)$ can be obtained by
 evaluating the change in the thermodynamic potential $\delta \Omega$ imposed by superconductivity~\cite{Abrikosov}
\begin{equation}
\Del \Omega=\int_0^\Delta \frac{d\lambda^{-1}_{\rm sc}}{d\Delta_1}\Delta_1^2d\Delta_1=-\frac{N_F\beta\Delta^4}{4},
\end{equation}
where $d\lambda_{\rm sc}^{-1}/d\Delta$ and $\Delta^2$ are defined by
 \req{eq:selfcons1}.
  For $\Delta^2$ we have
\be
\Delta^2 = \frac{1}{\beta} \left(\alpha (T) - \frac{2}{\lambda_{\rm sc} N_F}\right)= \frac{|\alpha^\prime|}{\beta} (T_c (M) - T).
 \label{n_9}
 \ee
The  change of the specific heat due to the superconducting ordering is $\Del C=-T\partial^2\delta \Omega/\partial T^2$. At $T=T_c(M)$, the specific heat  exhibits the jump given by
\begin{equation}
\frac{\Del C}{T_c}=\frac{3\gamma}{2\pi^2}\frac{[\alpha'(T_c (M))]^2}{\beta}
\label{eq:dC}
\end{equation}
where
 $\gamma=\pi^2N_F/3$ is the Sommerfeld  coefficient in  the metallic phase.

\begin{figure}
\centerline{\includegraphics[width=0.95\linewidth]{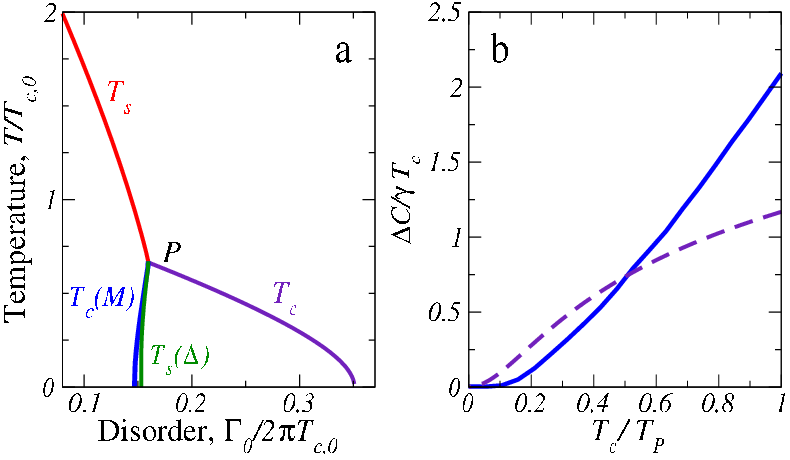}}
\caption{(Color online)
a) The phase diagram as a function of disorder, $\Gamma_0/2\pi \Tco$ for $\Gamma_\pi=0.2 \Gamma_0$ and $\Tso/\Tco=3$.
 Note that the region of the mixed phases gets narrower than the region for on-cite impurities, cf. \ref{fig:1}.  b)  The specific heat jump as a function of $T_c/T_P$. The solid line represents the specific heat jump $\Del C/T_c$ for normal--to--SC transition, and the dashed line shows $\Del C/T_c$ in the overdoped region for transition between SDW  and SC coexistence phases.
}
\label{fig:3}

\end{figure}

\begin{figure}
\centerline{\includegraphics[width=0.75\linewidth]{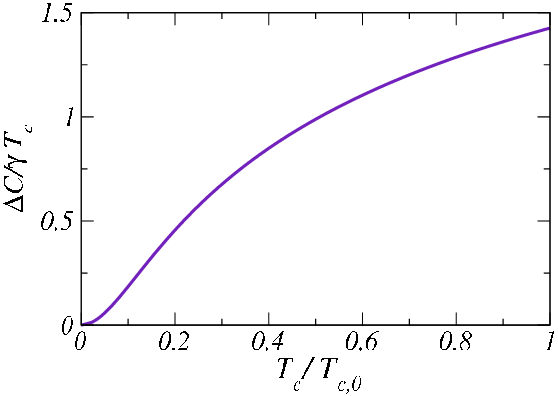}}
\caption{(Color online)
The jump $\Del C/T_c$ of the specific heat at a superconducting $T_c$
 without a competing SDW instability ($\lambda_{\rm sdw} =0$).
 $\Del C/T_c$ is plotted  as a function of $T_c/\Tco$, where
  $\Tco$ is the superconducting transition temperature in the clean limit.
}
\label{fig:4}
\end{figure}

The behavior of $\Del C/T_c$ as a function of doping--induced disorder is shown in Figs.~\ref{fig:1}b, \ref{fig:2}, and  \ref{fig:3}b. To evaluate $\Del C/T_c$ at the transition from a normal metal to a superconductor above optimal doping we use Eqs. (\ref{fr_all}) for $\alpha^\prime$ and $\beta$ in \eqref{eq:dC}. At large doping, when  $T_c$ is significantly suppressed and the system
 enters
 the regime of impurity--induced gapless superconductivity, specific heat jump $\Del C/T_c$ decreases with $T_c$
  as
   $\Del C/T_c \propto T^2_c$,  Ref.~\onlinecite{Kogan}.
Away from the gapless regime,
 the dependence of $\Del C/T_c$ on $T_c$ is more complex and differs from $T^2_c$, as the  dashed lines in Figs. ~\ref{fig:1}b, \ref{fig:2}, and  \ref{fig:3}b.    For completeness
 we present in Fig.~\ref{fig:4} $\Del C/T_c$ as a function of $T_c/T_{c,0}$ for an $s^\pm$ superconductor, when there is no competing SDW
  instability (i.e., when $\lambda_{\rm sdw} =0$) and the  $T_c$ line extends to $T_{c,0}$ in the clean limit.

For the transition from the preexisting SDW state into the mixed state below optimal doping, we compute  $\Del C/T_c$, \eqref{eq:dC}, using  $\alpha^\prime$ and $\beta$  from Eqs. (\ref{eq:alphamixed}) and (\ref{eq:beta}).  In this regime,
 SDW order strengthen as doping decreases,  and SDW correlations suppress both
 superconducting
   $T_c(M)$ and $\Del C/T_c$.  In particular, the rapid decrease of $\Del C/T_c$ at smaller dopings is
an indicator that fewer quasiparticle states participate in superconducting pairing, as the low-energy states are pushed away from the FS by SDW order.
We see therefore that $\Del C/T_c$ drops at deviations from optimal doping in both overdopped and underdopped regimes.

 It is essential that in the disorder model, the behavior of $\Del C/T_c$ in the underdoped and overdoped regimes  is governed by
   the
   single parameter $\Gamma_0 \propto n_{imp}$,
assuming that the ratio
     $\Gamma_\pi/\Gamma_0$ is kept constant.
      The same parameter $\Gamma_0$ also defines  $T_c/T_P$, where  $T_P=T_c (M \to 0)$ is the transition temperature at the tetracritical point.  One can therefore
 make a direct comparison with experiments by plotting $\Del C/T_c$ above and below optimal doping as the function of the same  $T_c/T_{P}$ using $\Del C$ defined by  \req{eq:dC}
 with
$\alpha'$ and $\beta$ given by \reqs{fr_all}
above optimal doping,  and
by
\reqs{eq:alphamixed} and \rref{eq:beta}  for a finite $M$ below optimal doping. Experiments show~\cite{budko_09,hardy_10,hardy_10a}
that $\Del C/T_c$ drops faster in underdoped regime, but in log-log plot the data from underdoped and overdoped regimes
 can be reasonably well fitted by a quadratic law $\Del C/T_c \propto T_c^2$.

We plot our $\Del C/T_c$  in Fig. \ref{fig:2} as  functions of $T_c/T_{P}$ in both linear and  log-log plots. We see that  $\Del C/T_c$ drops faster with decreasing $T_c$ in the underdoped regime, i.e. in the mixed phase. 
 This behavior is consistent with experimental  data. 
The $T$ dependence of $\Del C/T_c$ is not exactly $T^2$ , but looks reasonably close to $T^2$ in 
log-log plot 
 (right panel in Fig. \ref{fig:2}), although such plot puts more weight on the data at low $T_c$ where the $T-$dependence of 
$\Del C/T_c$ is the strongest. 

Finally, we note that $\Del C/T_c$ is discontinuous at the tetra-critical point $T_c = T_s = T_P$.
The discontinuity is the manifestation of
the discontinuous
 change in both $\alpha'$ and $\beta$ across $T_P$.
The discontinuity in $\alpha'$ is due to the fact that the second term in  (\ref{eq:alphamixed})
is zero
 for $M=0$, but is finite
 when $M \neq 0$
  and
contains temperature derivative of $(g^{(0)}_{\omega_n} -1) \propto M^2 \propto (T-T_P)$.
The discontinuity in $\beta$ is due to the feedback term $C^2/B$ in $\beta$, which  remains finite upon approaching $T_P$ from smaller dopings, 
 but is absent in the overdoped regime, where $T_s<T_c$.
  The interplay between discontinuities in $\alpha'$ and $\beta$ in the disorder model is such that
  $\Del C/T_c$ jumps up at $n_{\rm imp}$  once the system enters the mixed phase.

  The discontinuity in $\Del C/T_c$ at $T_P$ has also been found in the rigid band model.\cite{vvc_rigid} In that model,  however,
the magnitude and the sign of the jump in $\Del C/T_c$ depend on the FS geometry and
$\Del C/T_c$ may actually drop down upon entering into the  mixed phase.
We also emphasize that discontinuity in $\Del C/T_c$ only holds within the mean--field theory and gets rounded up and transforms into a maximum once
   we include fluctuations
  because then the thermodynamic average $\langle M^2\rangle $ is non-zero on both sides of the tetra-critical point.  This behavior of $\Del C/T_c$
in the presence of fluctuations is schematically illustrated in Fig.~\ref{fig:1}b by the dashed line.

\section{Conclusions\label{sec:6}} 
In this paper we obtained the phase diagram of doped Fe-pnictides and the specific heat jump $\Del C$ at the onset of superconductivity across the phase diagram under the assumption that doping introduces disorder but does not affect the band structure. The
  phase diagram is quite similar to the one obtained in the rigid band scenario and contains SDW and SC phases and the region where SDW and SC orders co-exist.  The ratio $\Del C/T_c$, which is a constant in a BCS superconductor, is non-monotonic across the phase diagram -- it has
   a maximum at the tetra-critical point at the onset of the mixed phase and drops at both larger and smaller dopings. The behavior
     at large and small dopings  is described in terms of the single parameter: the impurity density $n_{\rm imp}$.
    The non-monotonic behavior of  $\Del C/T_c$
     in the underdoped regime also holds in the rigid band model,\cite{vvc_rigid} but there the behavior of $\Del C/T_c$ at small and large dopings is generally uncorrelated.

We found  reasonably good agreement between our theory and  the experimental phase diagram of Ba(Fe$_{1-x}$Ru$_x$)$_2$As$_2$, in which Fe is subsituted by isovalent $Ru$ (see Refs.\cite{brouet,adam})
     and on the data for the doping dependence of the specific heat jump at $T_c$.~\cite{budko_09,hardy_10,hardy_10a}
This agreement  is a good indicator that our theory captures the key physics of non-monotonic behavior of $\Del C/T_c$, particularly the reduction of $\Del C/T_c$ in the mixed state. Whether the data can distinguish between rigid band and disorder scenarios is a more subtle issue as
   the interplay between doping-induced disorder and  doping-induced change in the band structure is likely to be material-dependent.
 Another subtle issue is the apparent $T^2$ dependence of the measured $\Del C/T_c$ on both sides of optimal doping.  Our log-log plots reproduce
  this dependence reasonably well, but our actual $\Del C/T_c$  are more mild than $T^2$.  One possible reason is our neglect of the doping dependence of $\gamma$ in the normal state specific heat $C = \gamma T$. In reality, $\gamma$ also decreases on both sides of optimal doping,\cite{hardy_10} and this should sharpen up the $T$ dependence  of $\Del C/T_c$.

Finally, there are certainly other elements of the physics of Ba(Fe$_{1-x}$Ru$_x$)$_2$As$_2$ which we neglected in our model. In particular,
 Brouet \textit{ et al.} demonstrated~\cite{brouet} that Fermi velocities in Ba(Fe$_{1-x}$Ru$_x$)$_2$As$_2$ are larger than in undoped BaFe$_2$As$_2$, this observation likely implies that electronic correlations are weaker in Ba(Fe$_{1-x}$Ru$_x$)$_2$As$_2$.
Dhaka \textit{ et al.} argued\cite{adam} that magnetic dilution due to Ru substitution contributes to the destruction of SDW order. These effects add on top of Ru-induced impurity scattering, which we studied here,  and call for more comprehensive analysis of isolvalent doping of Fe in 122 materials.

\begin{acknowledgements}
 We thank  S. Budko, P. Canfield, R. Fernandes, F. Hardy, I. Eremin, A. Kaminski, N. Ni,  J. Schmalian and A. Vorontsov for useful discussions. M.G.V. and A.V.C. are supported by NSF-DMR 0955500 and 0906953, respectively.
\end{acknowledgements}


\end{document}